# Neuronal Growth as Diffusion in an Effective Potential

Daniel J. Rizzo[1], James D. White[1,2], Elise Spedden[1], Matthew R. Wiens[1], David L. Kaplan[2], Timothy J. Atherton[1], Cristian Staii[1,*]

1. Department of Physics and Astronomy, Center for Nanoscopic Physics, Tufts University, Medford, MA 02155

2. Department of Biomedical Engineering, Tufts University, Medford, MA 02155

[*] Corresponding Author: Prof. C. Staii, E-mail: Cristian.Staii@tufts.edu

**Abstract**

Current understanding of neuronal growth is mostly qualitative, as the staggering number of physical and chemical guidance cues involved prohibit a fully quantitative description of axonal dynamics. We report on a general approach that describes axonal growth in vitro, on poly-D-lysine coated glass substrates, as diffusion in an effective external potential, representing the collective contribution of all causal influences on the growth cone. We use this approach to obtain effective growth rules that reveal an emergent regulatory mechanism for axonal pathfinding on these substrates.

The basic working unit of the nervous system is the neuron, a highly specialized cell consisting of three main structural components (Fig. 1a): the cell body (soma), several branching dendrites, and a single long axon that transmits electrical impulses to other neurons. In the developing nervous system each newly formed neuron extends an axon, which navigates through a complex and changing environment to reach dendrites from other target neurons and subsequently form functional connections called synapses [1]. Axonal guidance is governed primarily by the growth cone, a complex sensing unit located at the distal tip of the axon (Fig. 1a) that responds to a host of biophysical, chemical and mechanical cues. The high motility of the growth cone is based upon its cytoskeleton, a dynamic and flexible biopolymer network made from actin and microtubule filaments and their associated regulatory molecules. Collectively, these control the growth cone shape and its mechanical stability, mediate its sensing, guide the intracellular transport of various biomolecules, and direct axonal elongation [1].

It is now generally accepted that axons do not simply rely on an intrinsic "program of directions" that would uniquely specify each one of the billions of connections that form the neuronal architecture, being described instead by a set of general rules that apply across a large number of neurons and patterns of connections [2]. Finding the fundamental physical principles that govern the development of connections and communications between neurons is one of the key problems in biological physics. The main challenge faced when studying these processes *in vivo* lies both in the complex and highly controlled structure of neuronal matter in the nervous system, as well as in the complexity of the interplay between different environmental cues.

Therefore, an alternative approach [3, 4] is often used to uncover the basic rules that underlie the formation of functional neuronal connections. Within this approach one aims to create a simplified neuronal growth environment *in vitro* and to systematically investigate the effect that various cues: chemotactic, biochemical, mechanical, and topographical, have on the formation of neuronal networks. These studies show that physical stimuli (gradients of various molecular species, stiffness of the growth substrate, traction forces generated during axonal extension etc.) play a key role in the wiring of the nervous system [3-9]. However, our current understanding of neuronal growth is mostly qualitative, the vast complexity of the parameter space still prohibiting fully quantitative predictions of outcomes from given initial conditions such as: geometry of the neuronal circuit, type of biochemical cues on the growth substrate, topography or mechanical properties of the substrate.

A general description of axonal growth must take into account the inherent stochastic nature of this process due to different chemotactic signals, internal biochemical reactions and the randomness of external signals of various strengths [5, 7, 10-20]. Much of the previous work on stochastic effects has focused on describing axonal movement in the context of either intercellular diffusion of known neuron growth factors [5, 11-14], or intracellular events, such as polymerization of cytoskeletal structures [15-17], production of substrate adhesions sites [6], and transmembrane receptor activity [18]. In these cases, a model is motivated by a known underlying mechanism, whose validity is then tested against experimental data. Conversely, stochastic models have been recently introduced in order to provide purely phenomenological descriptions of axonal growth in some special cases, such as edge movement of the growth cone [10] or growth on asymmetrical surfaces [19].

In this paper, we present a general framework, based on the Fokker-Planck (FP) equation, to quantitatively describe and predict axon growth dynamics for cortical neurons (obtained from rat embryos) cultured on poly-D-lysine coated glass surfaces, through systematic measurements of axon growth velocity (throughout the paper referring specifically to the time-derivative of the axon arclength). We show that on these surfaces the axonal growth is governed by an effective potential, which incorporates all of the causal influences on the growth cone, and determines the evolution of its velocity distribution function. We find that axonal growth is not governed by simple diffusion, being instead described by a Laplace velocity distribution. The resulting time-dependent solutions of the 1-dimensional FP equation are used to extract an effective diffusion coefficient for axonal elongation rates. We demonstrate that this general model can be used to quantitatively describe the long-term behavior of growth cone dynamics and to predict experimental outcomes.

All measurements were performed on day 18 embryonic rat cortical neurons, cultured on poly-D-lysine (PDL) coated glass substrates. After plating, cells were allowed to incubate at 37°C for 8, 15, 19, 26, 33, or 46 hours. Samples are then removed from the incubator, loaded into the BioHeater™ Closed Fluid Cell, and imaged under bright-field using the optical stage of the MFP-3D-BIO Atomic Force Microscope (AFM, Asylum Research) [9, 19] (Fig. 1). The sample is then imaged every 5 minutes for a total period of 20 minutes (Fig. 2b,c). This time interval was chosen in order to allow enough time for typical axon growth to exceed the precision of our axon measurement (~0.1μm), while being short enough to accurately approximate an instantaneous velocity. The Nikon NIS-Elements Basic Research software [9, 19] is used to measure the change in axon length, yielding 3 consecutive values of the 5-minute average axon growth velocity. While Fig. 1 shows an example of axon extension, retraction (i.e. negative velocity) and zero-growth rates are also observed.

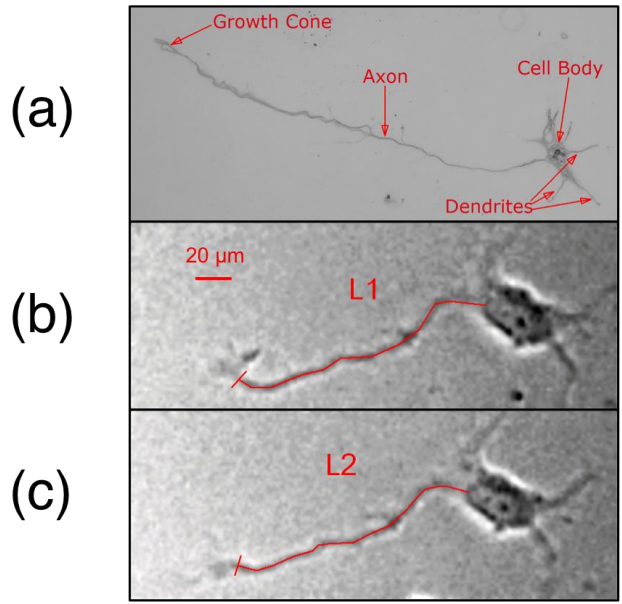

FIG. 1. (a) Cultured rat cortical neuron labeled with main structural components. (b), (c) Examples of axonal arclength measurements. Measurements are performed before (b) and after (c) a 5-minute wait period. The average velocity is determined from the difference between L1 and L2 divided by the 5-minute time interval. The growth description is presented as a one-dimensional probability distribution of the measured velocity.

Axon elongation rates are collected for each incubation period, resulting in 6 distinct time-lapse sets for the average growth velocity, each set containing between 140 and 250 data points. These are combined to form a single time-independent distribution containing a total of 986 observations (Fig. 2a). We note that the measured velocities follow a Laplace distribution, and not a Gaussian distribution that would be expected if the growth cone moved according to a Brownian random walk [21]. We show this by calculating the relative likelihood that observed data was produced by a Gaussian vs. Laplace distribution, i.e. the ratio of the products of the respective functions (Gaussian vs. Laplace) evaluated for all data points shown in Fig. 2a. The relative likelihood is found to be extremely small ($\sim 10^{-60}$) demonstrating that the Laplace distribution is definitively favored.

To gain a clearer insight into the growth dynamics that lead to the observed velocity distribution we construct a general model using the 1-dimensional Fokker-Planck equation:

$$\frac{\partial p(v,t)}{\partial t} = \frac{\partial}{\partial v}\left[\frac{\partial}{\partial v}\left(V(v)\right) \cdot p(v,t)\right] + D\frac{\partial^2 p(v,t)}{\partial v^2} \quad (1)$$

where $p(v,t)$ is the probability distribution for velocity, $V(v)$ is the effective potential that governs the evolution of this distribution, and $D$ is the diffusion coefficient in velocity space. We note that in general, equation (1) could also incorporate drift terms to account for biochemical or mechanical interactions between neurons, or between neurons and the growth substrate. For example, in our previous work we have used a constant velocity drift term to quantify axonal bias imparted by mechanical interactions between the growth cone and asymmetric growth surfaces [19]. The potential $V(v)$ in eqn. 1 can be obtained from the time-integrated data: first we find the stationary solution $p_s(v)$ by setting the time derivative to zero and then solve equation (1) for the potential $V(v)$:

$$\frac{V(v)}{D} = -\ln[p_s(v)] - \ln[\mathcal{N}] \quad (2)$$

where $N$ is a normalization constant. Hence, for a constant $D$ the effective potential that governs the axonal growth may be deduced directly from the observed time-independent distribution $p_s(v)$ as displayed in Fig. 2.

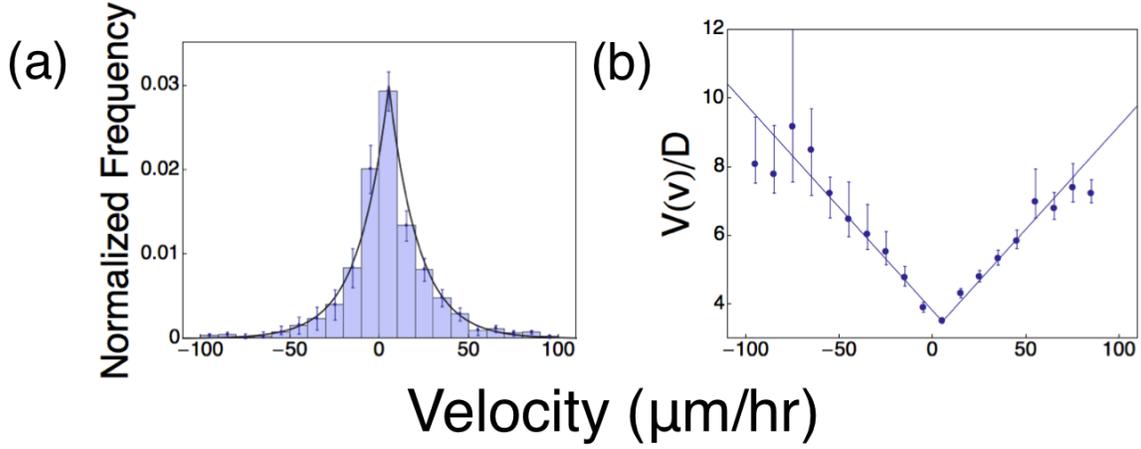

FIG. 2. (a) Time-integrated data of axon elongation rates. Fitting a Laplace distribution yields the mean velocity $v_c$ = 5.1 μm/hr and scale parameter $1/\kappa$ = 16.7 μm/hr. (b) V-shaped velocity-potential obtained from the probability distribution shown in (a) using equation (2). Error bars represent a 95% confidence interval of a binomial distribution.

The time-integrated data shown in Fig. 2a follows a Laplace distribution:

$$p_s(v) = \mathcal{N} \operatorname{Exp}(-\kappa |v - v_c|) \qquad (3)$$

with the parameter values: $\kappa$ = 0.06 hr/μm and $v_c$ = 5.1 μm/hr.

From eqns. (2) and (3) we obtain an effective potential:

$$\frac{V(v)}{D} = \kappa |v - v_c| - \ln[\mathcal{N}] \qquad (4)$$

The normalization constant *N* in Eq. 4, may be ignored, as it has no bearing on the final form of the normalized solution *p(v,t)*. We note that a simple random walk would result in a parabolic potential [21]. The rather uncommon V-shape of the effective potential provides insight into an effective mechanism governing axonal growth on PDL coated glass surfaces. The constant absolute value of the slope of the potential for all $v$, regardless of how far it may be from the preferred $v_c$, indicates that a constant effective drift force, corresponding to the derivative of the potential, "pushes" elongation rates toward the preferred value. That is, the axon appears to have a bimodal tendency to restore the growth velocity $v_c$ from "too fast" or "too slow" regimes. We speculate that this behavior may be a result of the previously observed bistable process involved in the growth of leading edge lamellipodia which act as the sensors of the growth cone [10].

We now turn to the time-evolution of the growth process. From the experimentally determined time-independent potential $V(v)/D$, one may solve for the time-dependent distribution $p(v,t)$ by substituting eqn. (4) into eqn. (1), and selecting an appropriate initial condition. A well-known method [21] is used to determine the time-dependent solution $p(v,t)$ by transforming (1) into a Schrödinger-like equation of the new function $q(v,t)$ governed by a new potential $V_s(v)$:

$$\frac{\partial q(v,t)}{\partial t} = -V_s(v)q(v,t) + D\frac{\partial^2 q(v,t)}{\partial v^2} \quad (5)$$

$$p(v,t) = \sqrt{p_s(v)}q(v,t) \quad (6)$$

$$V_s(v) = \frac{V'(v)}{4D} - \frac{V''(v)}{2} \quad (7)$$

Inserting (4) into (7), one finds that for a V-shaped potential, the corresponding Schrödinger potential will be $V_S(v) = D\kappa^2/4 + D\kappa\delta(v - v_c)$. Thus, one may obtain the solution of the Fokker-Planck equation for our system from the solution of the Schrödinger equation in the elementary delta-function potential well. The solution of the corresponding Schrödinger equation has a single bound state, $\psi_0(v) = \sqrt{\kappa/2}\exp[-\kappa|v - v_c|/2]$, and a continuum of eigenstates sorted by parity [21]:

$$\psi_k^{asym}(v) = \frac{1}{\sqrt{\pi}}\sin[k(v - v_c)] \tag{8}$$

$$\psi_k^{sym}(v) = \frac{2k\cos[k(v - v_c)] - \kappa\sin[k|v - v_c|]}{\sqrt{\pi(4k^2 + \kappa^2)}} \tag{9}$$

While the bound state has an eigenvalue of 0, the other eigenvalues are: $\lambda_k = D\kappa^2/4 + Dk^2$ where $k$ is a real number. Each unbound state has the typical exponential time-dependence: $\Psi_k(v,t) = \psi_k(v)\exp[-\lambda_k t]$. Given our choice of initial conditions (described below) and eqn. (6), the coefficients of the eigenfunction expansion are determined, which in turn yield the final solution upon summation and integration over $k$. Together, these determine the final solution of the FP equation (Fig. 3) [22].

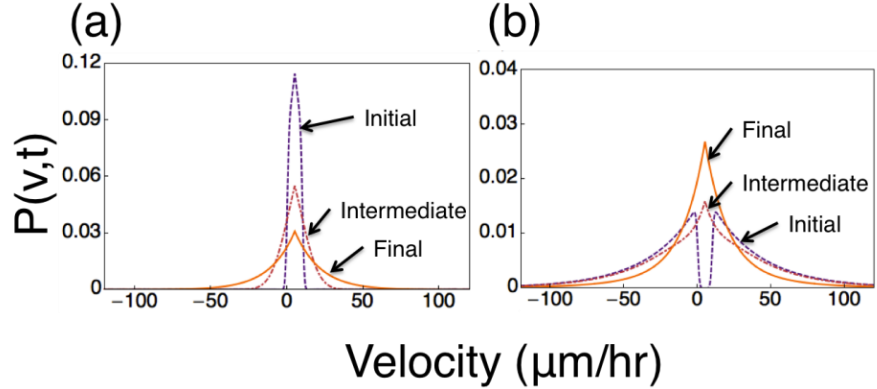

FIG. 3. Theoretical time-evolution of axonal velocity distributions that follow an initial truncated Laplace distribution (dotted line) defined in the range (a) [-2.5,12.5], and (b) (-∞, -2.5)∪(12.5, ∞). All units are in μm/hr. These initial distributions evolve in time toward the time-independent Laplace distribution given by eqn. (2) (represented by the continuous curves in (a) and (b), respectively). The time evolution is governed by the Fokker-Planck equation; two intermediate states are shown as dotted-dashed lines in (a) and (b). The diffusion coefficient for axonal growth velocity is determined by fitting these solutions to the corresponding initial subsets in the experimental time-lapse data (see Fig. 4).

Inspection of time-dependent data reveals that for all incubation periods (i.e. 8, 15, 19, 26, 33, or 46 hours), the velocity distribution is better described by the time-independent solution (3), than any time-dependent solution with physically reasonable initial conditions [22]. This shows that for each incubation period the overall velocity distributions are stationary over the timeframe of the observation (20 minutes). However, for any *individual* neuron, axon velocities might vary considerably among the 3 time points measured within the 20-minute observation timeframe, showing that measureable evolution does take place over a timescale of minutes. To quantify the dynamics on these shorter timescales, time-lapse velocity measurements for all neurons are combined to form two sets of 3 time-dependent distributions. One set incorporates only those neurons whose initial velocity falls in the range of [-2.5,12.5] μm/hr (Fig. 4b), and the other only those that do not fall in this range. In so doing, we limit our observation to two non-overlapping, non-equilibrium subpopulations of neurons, the evolution of which may be quantified with our solutions (Fig. 3). A best-fit value of $D$ for both experimental subsets is determined by selecting that $D$ which maximizes the likelihood of measuring the given data sets, while the FWHM of the peak of the likelihood is used to calculate an error on $D$ [22]. With this method, the joint fit of all incubation periods yields an effective value for the diffusion coefficient of $D = 1.2^{+0.4}_{-0.3} \times 10^4$ μm²/hr³.

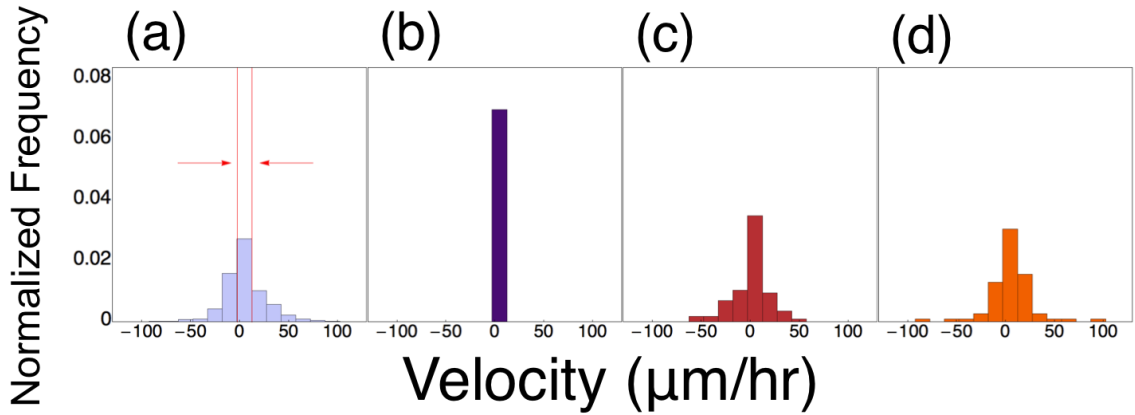

FIG. 4. Plot of initial measurements for all neurons from all incubation periods (a) before and (b) after limiting observation to specific subset of data, with velocities in the interval [-2.5,12.5] μm/hr, indicated with arrows. (c) The distribution of velocities for this subset of neurons measured after 5 and (d) 10 minutes, showing evolution toward a Laplace stationary distribution. Similar behavior is observed for the complimentary data set.

To our knowledge, this is the first instance in which a diffusion (or random motility) coefficient for growth cones has been measured. Although the above analysis outlines the measurement of velocity-space diffusion, we have also determined the position-space diffusion coefficient from axon length data. For regular diffusion, where the effective potential in the Fokker-Planck equation is $V(v) = \gamma v^2/2$ and the velocity distribution is Gaussian, the mean square displacement (MSD) grows linearly with time $\langle (x(t) - x_0)^2 \rangle = 2D_p t$, where the proportionality constant $D_p$ is the position diffusion (i.e. random motility) coefficient [23, 24]. We established, using a Monte Carlo simulation of an ensemble of random walkers where the Gaussian distribution was replaced with a Laplace distribution, that the linear time dependence of the MSD is expected to remain linear as well for the Laplace distribution. Using experimental data from all incubation periods, it was found that mean-squared axon length indeed has linear time dependence, with a diffusion coefficient: $D_p = (15 \pm 1)$ μm$^2$/hr [22].

A simple dimensional analysis argument demonstrates the consistency of the position and velocity space measurements. One may derive a characteristic position-space diffusion coefficient from the parameters in the Fokker-Planck equation $\kappa$ and $D$, by recognizing that these define a characteristic velocity of $v = 1/\kappa$ and a time scale of $\tau = 1/D\kappa^2$. Substituting experimental determined value of $\kappa$ and $D$, we have that: $D_p \sim v^2 \tau \sim 1/D\kappa^4$ has a value $D_p \sim 6$ μm$^2$/hr, which is comparable to the value obtained from the MSD analysis. We note that this value is also comparable to the diffusion coefficients measured for human peritoneal mesothelial cells *in vitro* [23], and about two orders of magnitude smaller than the values obtained for glioma cells, which are known to exhibit abnormally high motility rates [24].

In this paper, we have obtained effective rules for axonal growth on poly-D-lysine coated glass surfaces that extend beyond that of simple diffusion. Using the Fokker-Planck equation, we quantify the bimodal growth behavior of axons on these surfaces as diffusion in a V-shaped potential. This potential represents the collective contribution of all causal influences on the growth cone, and reveals an emergent growth rate regulatory mechanism for neurons on PDL coated surfaces. We emphasize that the symmetric V-shaped potential that governs the axonal growth rates on PDL coated glass surfaces, in the absence of other chemotactic sources is not universal, and does not describe neuronal growth in general. Other types of potentials are expected to be found on different growth environments and/or on different types of surfaces, such as asymmetric and/or spatially dependent potentials resulting from chemotactic sources, from growth on asymmetric or mechanically heterogeneous surfaces etc. Future studies that apply Fokker-Planck formalism (including possible drift terms and spatially dependent probability distributions) for studying growth dynamics under well-defined external conditions, such as neuron growth under controlled geometries on protein-patterned substrates, and on substrates of different stiffness or topography, may therefore provide a general framework for quantitative description of the role that different types of environmental cues have on axonal growth. Future studies might also provide new insight into the intracellular mechanisms that lead to the emergent growth rules observed in these controlled experiments, by correlating the growth potential with forces and stresses generated during growth (measured for e.g. by Traction Force Microscopy), as well as with cytoskeletal rearrangements measured using fluorescent techniques.

**Acknowledgement:** The authors thank Dr. Steve Moss's laboratory (Tufts Center for Neuroscience) for providing embryonic rat brain tissues. The authors gratefully acknowledge financial support for this work from National Science Foundation (NSF-CBET 1067093) and Tufts University.

**Author Contributions:** TJA, DLK, and CS planned and supervised the research. DJR, JDW, MW and ES cultured the cells and measured neuronal growth. DJR, TJA and CS analyzed the data and developed the theoretical model for axonal growth. All authors contributed to writing and revising the manuscript, and agreed on its final contents.

## References

[1]  L. A. Lowery and D. Van Vactor, Nat. Rev. Mol. Cell. Biol. **10**, 332 (2009).
[2]  H. D. Simpson, D. Mortimer, and G. J. Goodhill, Curr. Top. Dev. Biol. **87**, 1 (2009).
[3]  C. Staii, C. Viesselmann, J. Ballweg, J. C. Williams, E. W. Dent, S. N. Coppersmith, and M. A. Eriksson, Langmuir **27**, 233 (2011).
[4]  C. Staii, C. Viesselmann, J. Ballweg, L. Shi, G. Y. Liu, J. C. Williams, E. W. Dent, S. N. Coppersmith, and M. A. Eriksson, Biomaterials **30**, 3397 (2009).
[5]  G. J. Goodhill and J. S. Urbach, J. Neurobiol. **41**, 230 (1999).
[6]  M. Vanveen and J. Vanpelt, J. Theor. Biol. **159**, 1 (1992).
[7]  Y. E. Pearson, E. Castronovo, T. A. Lindsley, and D. A. Drew, Bull. Math. Biol. **73**, 2837 (2011).
[8]  D. Koch, W. J. Rosoff, J. Jiang, H. M. Geller, and J. S. Urbach, Biophys. J. **102**, 452 (2012).
[9]  E. Spedden, J. D. White, E. N. Naumova, D. L. Kaplan, and C. Staii, Biophys. J. **103**, 868 (2012).
[10] T. Betz, D. Lim, and J. A. Kas, Phys. Rev. Lett. **96**, 098103 (2006).
[11] G. J. Goodhill, Trends Neurosci. **21**, 226 (1998).
[12] G. J. Goodhill and H. Baier, Neural Comput. **10**, 521 (1998).
[13] H. G. Hentschel, and A. van Ooyen, Proc. Biol. Sci. **266**, 2231 (1999).
[14] R. Segev, and E. Ben-Jacob, Neural Networks **13**, 185 (2000).
[15] H. M. Buettner, Cell. Motil. Cytoskeleton **32**, 187 (1995).
[16] D. J. Odde, E. M. Tanaka, S. S. Hawkins, and H. M. Buettner, Biotechnol. Bioeng. **50**, 452 (1996).
[17] M. E. Robert and J. D. Sweeney, J. Theor. Biol. **188**, 277 (1997).
[18] H. Meinhardt, J Cell Sci **112**, 2867 (1999).
[19] R. Beighley, E. Spedden, K. Sekeroglu, T. Atherton, M. C. Demirel, and C. Staii, Appl. Phys. Lett. **101**, 143701 (2012).
[20] A. Granato and J. Van Pelt, Dev. Brain Res. **142**, 223 (2003).
[21] H. Risken, *The Fokker-Planck Equation: Methods of Solution and Applications* (Springer, Berlin, 1996), 2$^{nd}$ ed.
[22] See Supplemental Material at [URL inserted by publisher] for further explanation of solutions, fitting and simulation.
[23] P. K. Maini, D. L. McElwain, and D. I. Leavesley, Tissue Eng. **10**, 475 (2004).

[24]   K. R. Swanson, Math. Comput. Model. **47**, 638 (2008).

**Supplemental Material for**

*Neuronal Growth as Diffusion in an Effective Potential*

Daniel J. Rizzo[1], James D. White[1,2], Elise Spedden[1], Matthew R. Wiens[1], David L. Kaplan[2], Timothy J. Atherton[1], Cristian Staii[1,*]

1. Department of Physics and Astronomy, Center for Nanoscopic Physics, Tufts University, Medford, MA 02155
2. Department of Biomedical Engineering, Tufts University, Medford, MA 02155

[*] Corresponding Author: Prof. C. Staii, E-mail: Cristian.Staii@tufts.edu

In this supplemental material, we outline the calculation of the time-dependent solutions of the Fokker-Planck equation for a V-shaped potential (S1), and describe how this solution is used to find a best-fit value for the velocity-space diffusion coefficient (S2). We then present simulation results that demonstrate the linear behavior of mean square displacement (MSD) with time for random-walkers whose velocities are drawn from a Laplace distribution. These are presented alongside experimental time-dependent MSD data (S3). Finally, we present experimental data that shows the general behavior of axon velocities that lead to Laplace distributions for each incubation period (S4).

**S1. Time-Dependent Fokker-Planck Solution in a V-shaped Potential**

Fig 2a (main text) shows that the time-independent experimental data is well described by a Laplace distribution. From this, we have derived a V-shaped potential using the Fokker-Planck Equation (Fig. 2b). Given that the shape of the potential is known, a time-dependent solution can be determined. To do this, a well-known method is used that transforms the Fokker-Planck equation of the function $p(v,t)$ into a Schrodinger-like equation of the function $q(v,t)$ [1]. The relationship between these functions is:

$$p(v,t) = \sqrt{p_s(v)} q(v,t) \tag{SE1}$$

where $p_s(v)$ is the time-independent distribution given by equation (3) in the main text. As stated in the main text, solving for $p(v,t)$ in a V-shaped potential is equivalent to solving for $q(v,t)$ in an elementary Dirac Delta function potential well, using the transformation (SE1). Thus, the eigenstates of $q(v,t)$ are:

$$\psi_0(v) = \sqrt{\frac{\kappa}{2}} \operatorname{Exp}\left[\frac{-\kappa |v - v_c|}{2}\right] \tag{SE2}$$

$$\psi_k^{asym}(v) = \frac{1}{\sqrt{\pi}} \sin[k(v - v_c)] \tag{SE3}$$

$$\psi_k^{sym}(v) = \frac{2k \cos[k(v - v_c)] - \kappa \sin[k|v - v_c|]}{\sqrt{\pi(4k^2 + \kappa^2)}} \tag{SE4}$$

where (SE2) has an eigenvalue $\lambda=0$, and the states (SE3) and (SE4) have a continuum of real eigenvalues defined as:

$$\lambda_k = \frac{D\kappa^2}{4} + Dk^2 \qquad (SE5)$$

where *k* is a real number.

These solutions define the usual time-dependence:

$$\Psi_k(v,t) = \psi_k(v)\text{Exp}[-\lambda_k t] \qquad (SE6)$$

As explained in the main text, the two non-intersecting, non-equilibrium initial conditions for $p(v,t)$, with the corresponding initial $q(v,t)$ eigenfunctions (see SE1) are given by:

$$p_1(v,0) = \begin{cases} 0 & : \quad v<0, v>10 \\ \frac{\kappa}{2}\text{Exp}[\kappa|v-v_c|] & : \quad 0 \leq v \leq 10 \end{cases} \qquad (SE7)$$

$$q_1(v,0) = \frac{p_1(v,0)}{\sqrt{p_S(v)}} = \begin{cases} 0 & : \quad v<0, v>10 \\ \sqrt{\frac{\kappa}{2}}\text{Exp}[\kappa|v-v_c|/2] & : \quad 0 \leq v \leq 10 \end{cases} \qquad (SE8)$$

$$p_2(v,0) = \begin{cases} \frac{\kappa}{2}\text{Exp}[\kappa|v-v_c|] & : \quad v<0, v>10 \\ 0 & : \quad 0 \leq v \leq 10 \end{cases} \qquad (SE9)$$

$$q_2(v,0) = \frac{p_2(v,0)}{\sqrt{p_S(v)}} = \begin{cases} \sqrt{\frac{\kappa}{2}}\text{Exp}\left[\frac{\kappa|v-v_c|}{2}\right] & : \quad v<0, v>10 \\ 0 & : \quad 0 \leq v \leq 10 \end{cases} \qquad (SE10)$$

In order to determine the final solution, the coefficients for the eigenfunction expansion must be determined for each state. This is done by projecting each set of eigenstates onto the respective initial conditions, where the subscript *i* =1, 2 denotes which of the two initial conditions are being used:

$$C_{0,i} = \int_{-\infty}^{\infty} q_i(v,0)\psi_0(v)dv \qquad (SE10)$$

$$C_{k,i}^{asym} = \int_{-\infty}^{\infty} q_i(v,0)\psi_k^{asym}(v)dv \qquad (SE11)$$

$$C_{k,i}^{sym} = \int_{-\infty}^{\infty} q_i(v,0)\psi_k^{sym}(v)dv \qquad (SE12)$$

Thus, the final form of the function $q(v,t)$ can be determined by integrating over all continuous eigenstates, and adding the one discrete state $\psi_0$:

$$q_i(v,t) = C_{0,i}\psi_0(v) + \int_0^{\infty} \left( C_{k,i}^{sym}\psi_k^{sym}(v) + C_{k,i}^{sym}\psi_k^{sym}(v) \right) \text{Exp}[-\lambda_k t]dk \qquad (SE13)$$

The final solution for $p(v,t)$ is determined by plugging (SE13) into (SE1) (with $p_s(v)$ given by equation (3) in the main text).

## S2. Maximum Likelihood Fitting Procedure and Errors

For $N$ measured velocities $v_i$, each with known associated time $t_i$ and probability distribution $p(v,t)$, the likelihood function is defined as:

$$L(D) = \prod_{i=1}^{N} p(v_i, Dt_i)$$

The value of the diffusion coefficient $D$ that maximizes $L$ is the best-fit value. The reported error on this value is the FWHM of the peak in the likelihood function. Here, the natural log of the likelihood function is presented for all data combined, and the ln[2] is subtracted from the maximum to find the FWHM.

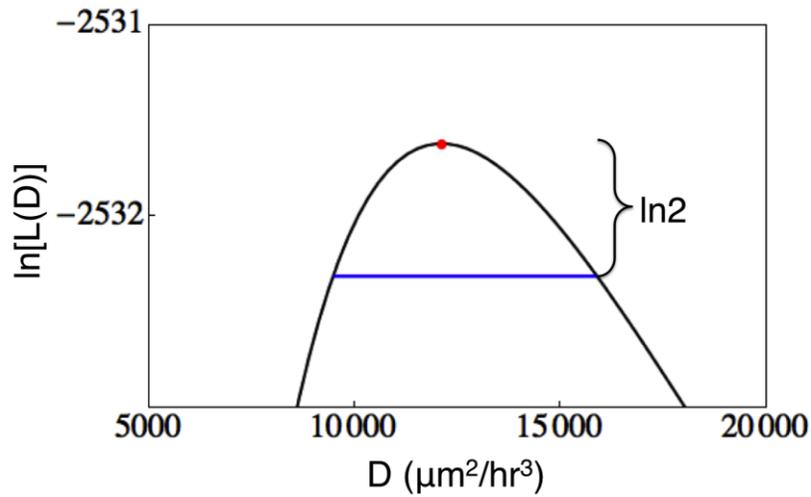

FIG. S1. Maximum likelihood fitting for $D$.

## S3. Determination of MSD with time for Laplace Velocity Distribution

We have performed a Monte-Carlo simulation of an ensemble of random walkers whose velocities are drawn from both Gaussian and Laplace velocity distributions for a given diffusion coefficient, and distribution width.

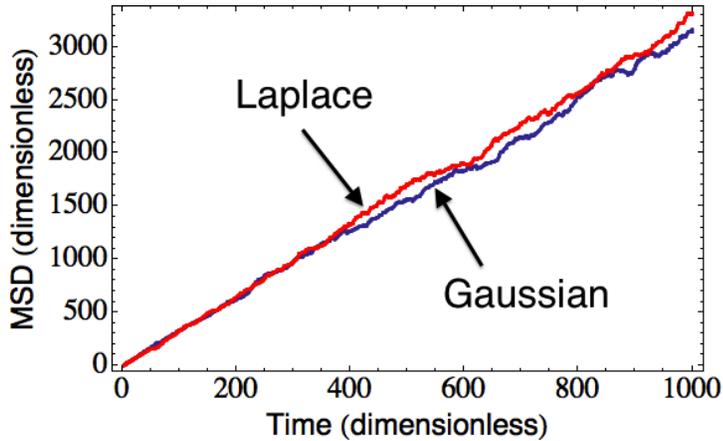

FIG. S2. Monte Carlo simulation of mean square displacements for random walkers drawn from Gaussian (blue) and Laplace (red) distributions.

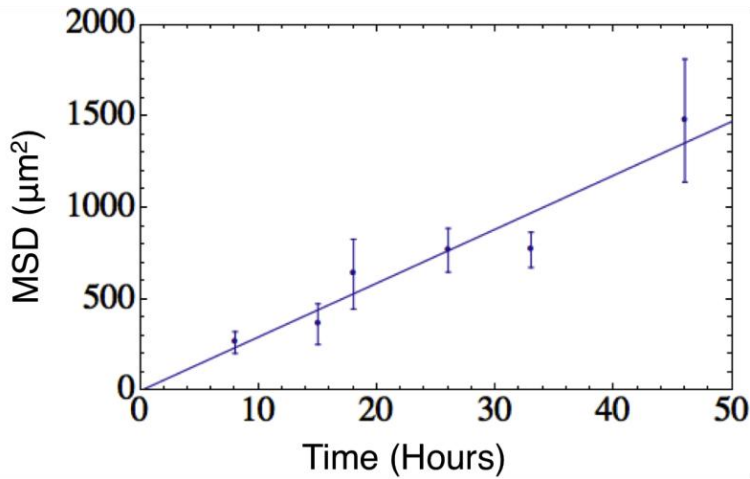

FIG. S3. Experimentally determined mean square displacement vs. time for the ensemble of growth cones. The error bars represent the standard error on the MSD for each time point.

The slope for the Gaussian MSD is equal to $2D_p$ (see main text). Given the observed similarity in slope between Laplace and Gaussian cases (Fig. S2), the experimental value of $D_p$ is extracted from the MSD assuming a Gaussian random walk. From the slope of the linear fit in Fig. S3 we obtain: $D_p = (15\pm1)$ μm$^2$/hr, where the reported uncertainty is the error in the linear fit.

## S4. Time-Dependent Distributions by Incubation Period

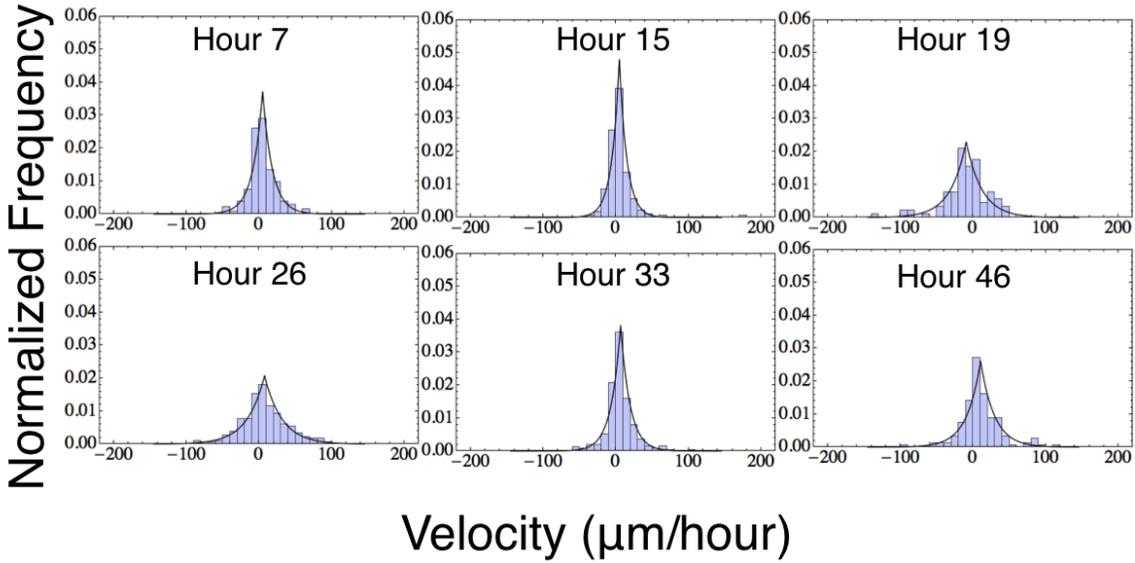

| Incubation Period (hr) | Mean/Minimum (μm/hr) | Slope of V(v)/D (hr/μm) |
|---|---|---|
| 8 | 4.7 | 0.08 |
| 15 | 4.5 | 0.10 |
| 19 | -9.5 | 0.05 |
| 26 | 7.3 | 0.04 |
| 33 | 6.3 | 0.08 |
| 46 | 10.3 | 0.05 |

FIG. S4. *Top*: velocity distributions at different incubation periods. *Bottom*: table showing the velocities at which the V-shape potential has a minimum, and the gradient (slope) of the potential for each incubation period.

When attempting to fit a time-dependent solution to the data sets shown in Fig. S4 it was determined that they were better described by the time-independent solution then any time-dependent solution. To quantify this, the initial condition was assumed to be δ($v$), which is a physically reasonable assumption given that no axons are growing during plating. Given this condition, the likelihood (example shown below) as a function of the fit parameter $Dt$ (i.e. the product between diffusion coefficient and time) is observed to plateau at high values of $Dt$, without the presence of a local maximum (i.e. best-fit parameter $Dt$). This can be explained by noting that the time-dependent solution approaches the time-independent solution for large $Dt$, and thus the plateau is the result of the time-dependent solution effectively reaching the equilibrium or stationary state. The fact that the likelihood is at its highest value at this point reveals that this stationary solution is a better fit than any previous time point, i.e. for each incubation period the overall velocity distributions are stationary over the timeframe of the observation (20 minutes).

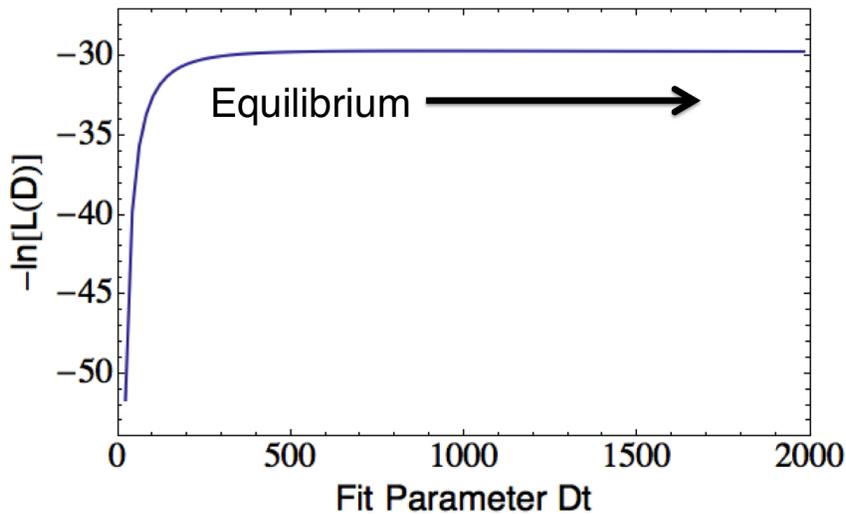

[1]   H. Risken, *The Fokker-Planck Equation: Methods of Solution and Applications* (Springer, Berlin, 1996), 2nd edn.